\documentclass[11pt]{iopart}
\usepackage{psfrag,graphicx,iopams,amstext,setstack}
\begin{document}
\title{Current fluctuations in stochastic systems with long-range memory}
\author{R. J. Harris and H. Touchette}
\address{School of Mathematical Sciences, Queen Mary University of London, Mile End Road, London, E1 4NS, United Kingdom.}
\eads{\mailto{rosemary.harris@qmul.ac.uk}, \mailto{h.touchette@qmul.ac.uk}}

%\newcommand{\corr}{\textbf}
% To find the corrections, search for ``corr''

\begin{abstract}
We propose a method to calculate the large deviations of current fluctuations in a class of stochastic particle systems with history-dependent rates.  Long-range temporal correlations are seen to alter the speed of the large deviation function in analogy with long-range spatial correlations in equilibrium systems.  We give some illuminating examples and discuss the applicability of the Gallavotti-Cohen fluctuation theorem.
\end{abstract}
\pacs{02.50.-r, 05.40.-a, 05.40.Fb}

\section{Introduction}
\label{s:intro}

Within the context of stochastic many-particle systems there has recently been considerable interest in the calculation of distributions of current fluctuations, see e.g.,~\cite{Derrida07b}.  Typically such distributions obey a large deviation principle with a rate function playing a role analogous to the entropy in equilibrium systems.\footnote{See~\cite{Touchette09b} for a general discussion of the large deviation approach to statistical mechanics.}.  Under rather general conditions, the ratio of probabilities for ``forward'' and ``backward'' particle currents obeys a so-called ``fluctuation theorem''~\cite{Evans93,Gallavotti95} as reflected in a particular symmetry property of the large deviation function, see e.g.,~\cite{Lebowitz99,Me07}.

The majority of previous works on stochastic dynamics have been concerned with Markovian systems.  However, the inherent memoryless assumption may be an inappropriate approximation for the modelling of many ``real-life'' processes where long-range temporal correlations are known to be important, see e.g.,~\cite{Rangarajan03,Mandelbrot99,benAvraham00,Mantegna99} and references therein.  On a coarse-grained scale, memory effects have been successfully treated using generalized Langevin or Fokker-Planck equations~\cite{Hanggi78,Volkov83,Metzler99}.   At the microscopic level of modelling a well-established paradigm 
is the continuous-time random walk~\cite{Montroll65}; for some biological applications see~\cite{Murray02}.  The validity of fluctuation relations for non-Markovian systems has also been explored using both these approaches~\cite{Mai07,Speck07b,Ohkuma07,Esposito08,Andrieux08d,Chechkin09}. 

An alternative strategy to introduce long-range memory into microscopic dynamics is to introduce an explicit history-dependence in the rates as is done, for example, in the ``elephant'' and ``Alzheimer'' walks~\cite{Hod04,Schutz04,daSilva05,Paraan06,Cressoni07,Kenkre07} or the negative mobility random walk of Cleuren and Van den Broeck~\cite{Cleuren03}.  Motivated by such models, in the present contribution we consider current large deviations in a class of stochastic processes where the rates at a given time depend on the past history of the current.  For such systems we propose a ``temporal additivity principle'' as a method for calculation of rate functions.   Our approach is in the spirit of the spatial additivity principle introduced by Bodineau and Derrida~\cite{Bodineau04}; we demonstrate it on various random-walk models and discuss its wider applicability to many-particle systems.  The examples considered allow us to illuminate some generic effects of the long-range memory (in particular, a modified ``speed'' in the large deviation principle) and test the validity of the fluctuation theorem.

\section{Stochastic framework}

Let us first consider a continuous-time \emph{Markov} process in discrete state space.  A particular realization of its history over the time interval $[0,t]$ is given by $\{\sigma(\tau),0\leq\tau\leq t\}$.  We assume here time-independent dynamics which are specified by the transition rates $w_{\sigma'\sigma}$; in other words, the probability for a transition from state $\sigma$ to state $\sigma'$ in the infinitesimal time interval $[t,t+\rmd t]$ is $w_{\sigma'\sigma} \, \rmd t$.

In the following we use script letters to denote functionals of the trajectory $\sigma(\tau)$.
In particular, we define a counter $\mathcal{Q}_t$ which has initial condition $\mathcal{Q}_0=0$ and increases by an amount $\Theta_{\sigma'\sigma}$ each time the transition $\sigma \to \sigma'$ occurs.  If the matrix $\Theta$ is real and antisymmetric we can say that it describes a real or abstract ``current'' through the system.  For example, the time-integrated particle current from left to right, say, is given by adding 1 every time a particle hops to the right and subtracting 1 every time a particle hops to the left.  The fluctuations and symmetry properties of general particle currents are central to our investigations.

The integrated current $\mathcal{Q}_t$ is time-extensive and it follows that $\mathcal{Q}_t/t$ is a time-intensive functional.  We denote a given realization of the latter's history by $\{q(\tau),0\leq\tau\leq t\}$ and use the symbol $j$ for the terminal value, i.e., $q(t)=j$.  Although $\mathcal{Q}_t$ is discrete, $\mathcal{Q}_t/t$ approaches a continuous random variable as $t \to \infty$ and we assume it satisfies a large deviation principle with ``rate function'' $I$ and ``speed'' $t$, i.e.,
\begin{equation}
\text{Prob}(\mathcal{Q}_t/t=j) \sim \rme^{-t I_w(j)}, \label{e:ldf}
\end{equation}
where the symbol $\sim$ means logarithmic equality in the limit of large $t$.\footnote{The term ``speed'' is used in large deviation theory to denote the $t$-dependent coefficient in front of the rate function \cite{Ellis95,Touchette09b}; it has nothing to do with the physical speed.}
Here the precise form of $I_w$ depends on the particular process considered, i.e., on the set of rates $w_{\sigma'\sigma}$.  As a concrete example the reader is invited to think of a single particle on a one-dimensional (1D) lattice which hops to the right with rate $v_R$.  The integrated particle current in time $t$ is simply the number of jumps made by the particle and the time-averaged (intensive) current obeys a large-deviation principle with rate function $I_{v_R}(j)=v_R-j+j\ln(j/{v_R})$ which is simply obtained from considering the asymptotic properties of the Poisson process.  Large deviation functions have also been investigated for more complicated systems of many interacting particles such as the exclusion process~\cite{Derrida04b,Bodineau06} and the zero-range process~\cite{Wijland04,Me05,Me06b}.  The latter model also illustrates some interesting subtleties which may arise when the state space is unbounded.  In particular, for ergodic systems with finite state space the rate function is independent of the initial state of the system (i.e., the initial particle configuration) but this may not be true for infinite state space \cite{Dembo98,Touchette09b}.  We defer consideration of such issues to a later publication and restrict ourselves here to situations where there is no initial-configuration dependence.

We now generalize the previous setup to consider a simple type of \emph{non-Markovian} process where the rates $w_{\sigma'\sigma}$ depend explicitly on $\sigma$, $\sigma'$ \emph{and} the value of $\mathcal{Q}_t/t$.  This introduces a dependence on the complete history of the process although we note that, in the enlarged current-configuration state space, the process \emph{is} Markovian.  This class of systems includes analogues of the ``elephant'' random walk introduced as a toy non-Markovian model in~\cite{Schutz04} and further analysed in~\cite{daSilva05,Paraan06,Kenkre07}.
In order to avoid singularities which arise at $\tau=0$ we assume that our observations start at time $t_0$ where $0 \ll t_0 \ll t$.  Physically, non-zero $t_0$ allows for initial ``transient'' behaviour with different transition rates; this is analogous to different probabilities at the first timestep in the original discrete-time elephant~\cite{Schutz04}.

Systems of the type defined above are expected to show long-range memory effects which threaten the simple form of the current large deviation principle~(\ref{e:ldf}).    In the next section we propose a general principle to obtain the asymptotic distribution of current fluctuations in such models.

\section{Temporal additivity principle}

Here we outline the steps leading to a ``temporal additivity principle''~(\ref{e:result}) noting that many elements of the argument are analogous to those given for spatial additivity in~\cite{Bodineau04}.

First we divide the time interval $[t_0,t]$ into $N$ sub-intervals of size $\Delta \tau$ and write $t_n=t_0+n\Delta\tau$ with $t_N=t$.  We use the notation $p(\sigma_n,q_n,t_n)$ for the probability that the system is found in the state $(\sigma_n,q_n)$ at time $t$ and $p(q_{n+1},\sigma_{n+1},t_{n+1}|q_n,\sigma_n,t_n)$ for the two-point conditional probability.  Since the system is Markovian in the joint state-current space, we have the Chapman-Kolmogorov equation
\begin{eqnarray}
\fl p(q_N,\sigma_N,t|q_0,\sigma_0,t_0) \nonumber \\
\fl = \underset{\sigma_1,\ldots, \sigma_{N-1}}{\sum_{q_1, \ldots, q_{N-1}}} p(q_N,\sigma_N,t|q_{N-1},\sigma_{N-1},t_{N-1})\cdots p(q_2,\sigma_2,t_2|q_1,\sigma_1,t_1)p(q_1,\sigma_1,t_1|q_0,\sigma_0,t_0) \label{e:CK}
\end{eqnarray}
Now if $\Delta \tau \gg 0$, then $p(q_{n+1},\sigma_{n+1},t_{n+1}|q_n,\sigma_n,t_n)$ can be assumed independent of $\sigma_n$, at least for an ergodic system with finite state space.   In this case, summing over the $\sigma_n$'s in~(\ref{e:CK}) gives
\begin{equation}
p(q_N,t|q_0,t_0) = \sum_{q_1, \ldots, q_{N-1}} p(q_N,t|q_{N-1},t_{N-1})\cdots p(q_2,t_2|q_1,t_1)p(q_1,t_1|q_0,t_0) \label{e:CK2}
\end{equation}
This shows that, when considering long time slices, the dynamics in the $q$-space is Markovian, although we emphasize that the underlying dynamics of the system's configurations (i.e., the dynamics in $\sigma$-space) is non-Markovian.

Next we take $t$ and $N$ large, whilst preserving their ratio, so that $t \gg \Delta \tau$ meaning that $q(\tau)$ is almost constant during each time slice (adiabatic or quasistatic approximation).  The observed current averaged over the time interval $(t_n,t_{n+1}]$ is given by
\begin{equation}
q^{(n)}_{\Delta\tau}=\frac{q_{n+1}t_{n+1}-q_n t_n}{\Delta\tau} 
\end{equation}
and thus, using the large deviation principle of (\ref{e:ldf}), we have
\begin{equation}
p(q_{n+1},t_{n+1}|q_n,t_n) \approx A_n \rme^{-\Delta\tau I_{w(q_n)}(q^{(n)}_{\Delta\tau})} \label{e:slice}
\end{equation}
where $A_n$ is a $q$-independent constant.  Here the subscript on the Markovian rate function $I$ emphasizes that it depends on the value of $q_n$ via the dependence on the (assumed) constant rates $w_{\sigma'\sigma}(q)$.  Substituting~(\ref{e:slice}) in~(\ref{e:CK2}) yields
\begin{equation}
p(q_N,t|q_0,t_0) \approx A \sum_{q_1, \ldots, q_{N-1}} \rme^{-\sum_{n=0}^{N-1} \Delta\tau I_{w(q_n)}(q^{(n)}_{\Delta\tau})}. \label{e:above}
\end{equation}

At this point, we pass to the continuum limit by taking $t, \Delta \tau \to \infty$ with $\Delta\tau/t\to 0$, in which case $N\to\infty$.  In this limit $q_n$ becomes a continuous path, so that $q_n \to q(\tau)$ and $q^{(n)}_{\Delta\tau} \to q(\tau) + \tau q'(\tau)$.  Then the sum in~(\ref{e:above}) becomes a path integral over the trajectories $q(\tau)$ and (using $j_0$ and $j$ for the fixed boundary values) we have
\begin{equation}
p(j,t|j_0,t_0) \sim \int_{q(t_0)=j_0}^{q(t)=j} \mathcal{D}[q]\,  \rme^{-\int_{t_0}^t I_{w(q)}(q+\tau q')\,\rmd\tau}.
\end{equation}
In the $t \to \infty$ limit, we expect this path integral to be dominated by the path in $q$-space for which the exponent in the integral is least negative, and so we apply a ``saddle-point approximation'' to arrive at our central result for the asymptotic distribution of $\mathcal{Q}_t/t$:
\begin{equation}
\text{Prob}(\mathcal{Q}_t/t=j) \sim \exp \left[- \min_{q(\tau)}\int_{t_0}^t I_{w(q)}(q+\tau q') \, \rmd\tau \right].
\label{e:result}
\end{equation}
The integral above has to be minimized over all trajectories $q(\tau)$ with $q(t_0)=j_0$ and $q(t)=j$.  Note that for notational brevity we have suppressed the dependence on $q_0, t_0$ on the left-hand side of this equation in anticipation that the initial condition will not play an important role.  Finally, to make the dependence on $t$ more explicit, we can re-write the result of (\ref{e:result}) as
\begin{equation}
\text{Prob}(\mathcal{Q}_t/t=j)\sim \rme^{-t^\alpha E(j)},
\end{equation}
where $\alpha$ is some positive constant and
\begin{equation}
E(j) = \lim_{t \to \infty} \min_{q(\tau)} \frac{1}{t^\alpha} \int_{t_0}^t I_{w(q)}(q+\tau q') \, \rmd\tau.
\end{equation}
If this function exists and is not everywhere zero, then we have that $\mathcal{Q}_t/t$ satisfies a large deviation principle with speed $t^\alpha$ and rate function $E(j)$. For Markovian processes, the exponent $\alpha$ is usually equal to 1 \cite{Dembo98,Touchette09b}.  However, we shall see in the next section that the $q$-dependence of the rates in our setup can give rise to a non-standard speed with $\alpha\neq 1$.

\section{Random-walk examples}
\label{s:rw}

As a simple pedagogical model, we first consider a ``uni-directional random walk'' on a 1D lattice.  Here $q(\tau)$ denotes the velocity trajectory (number of jumps divided by time) of a single particle which hops at time $\tau$ from site $i$ to site $i+1$ with rate $v_R(q)$.   In this case, since the associated Markovian process is Poissonian, we have
\begin{equation}
I_{v_R(q)}(q+\tau q') = v_R(q)-(q+\tau q')+(q+\tau q')\ln\left(\frac{(q+\tau q')}{v_R(q)}\right) \quad \text{for $q>0$}
\end{equation}
and minimizing the integral in~(\ref{e:result}) leads to the Euler-Lagrange equation for the minimum path
\begin{equation}
\frac{dv_R}{dq}-q \frac{dv_R/dq}{v_R} - \frac{2 \tau q'}{q+\tau q'} - \frac{\tau^2 q''}{q+\tau q'}=0. \label{e:ELv}
\end{equation}

To explore scenarios where the rate for the particle to move is directly proportional to the average velocity up to that instant, we set $v_R(q)=a q$.   In this case~(\ref{e:ELv}) is a linear differential equation which has general solution
\begin{equation}
q(\tau)=C_1 \tau^{-1} + C_2 \tau^{a-1}.
\end{equation}
The boundary conditions $q(t_0)=j_0$ and $q(t)=j$ fix the constants as
\begin{equation}
C_1=\frac{j_0 t_0 t^a - j t_0^a t}{t^a-t_0^a}, \qquad C_2=\frac{j t - j_0 t_0 }{t^a-t_0^a},
\end{equation}
and, integrating over this path, it is straightforward but tedious to show that the asymptotic behaviour for large $t$ is
\begin{equation}
\min_{q(\tau)}\int_{t_0}^t I_{v(q)}(q+\tau q') \, \rmd\tau \sim
\left\{
\begin{array}{ll}
j t_0^a t^{1-a} & \text{for $a < 1$} \\
(a-1)j_0 t_0 \ln t & \text{for $a > 1$}.
\end{array}
\right.
\end{equation}
Physically, we note that at $a=1$ there is a crossover between an ``escape'' regime and a ``localized'' regime.  For $a>1$ the mean velocity of the particle grows indefinitely with time, there is no stationary state and no large-deviation principle.  On the other hand, for $a<1$ the system approaches a state with zero velocity and we have
\begin{equation}
\text{Prob}(\mathcal{Q}_t/t=j) \sim \rme^{-t^{1-a} j t_0^a}, \qquad \text{for $j >0$}. \label{e:exp}
\end{equation}
i.e., a large deviation principle with speed $t^{1-a}$ and rate function $j t_0^a$.  
In fact, this is an exponential distribution for $\mathcal{Q}_t/t$ so one straightforwardly finds the asymptotic moments
\begin{equation}
E[ \mathcal{Q}_t/t ] \sim \frac{1}{t_0^a t^{1-a}}, \qquad \text{Var}[\mathcal{Q}_t/t] \sim \frac{1}{t_0^{2a} t^{2(1-a)}}.  
\end{equation}
From the latter equation one trivially sees that $\text{Var}[\mathcal{Q}_t] \sim (t/t_0)^{2a}$ and hence that there is a transition between ``subdiffusive'' and ``superdiffusive'' behaviour at $a=1/2$.  This is related to a similar transition found in the elephant random walk of Sch\"utz and Trimper~\cite{Schutz04}, a connection to which we shall later return.

Next we turn to the more interesting case of a bi-directional random walk, again on a 1D lattice. For full generality, we now count separately the total number of jumps to the right, $\mathcal{Q}_{+t}$, and the total number of jumps to the left, $\mathcal{Q}_{-t}$.  The net integrated current is given by $\mathcal{Q}_t=\mathcal{Q}_{+t}-\mathcal{Q}_{-t}$ and we choose rates which can depend on the present values of both $\mathcal{Q}_{+t}/t$ and $\mathcal{Q}_{-t}/t$, i.e., at a given point in the trajectory the rates are generically $v_R(q_+,q_-)$ and $v_L(q_+,q_-)$ for jumps to the right and left respectively.   In an obvious generalization of the previous treatment we then have
\begin{eqnarray}
\fl \text{Prob}(\mathcal{Q}_t/t=j) \sim  \exp\left\{
\min_{q+(\tau),q-(\tau)} \int_{t_0}^t \left[ v_R - (q_+ + \tau q_+') + (q_+ + \tau q_+')\ln\left(\frac{q_+ + \tau q_+'}{v_R}\right)\right.\right. \nonumber \\ \left.\left. + v_L - (q_- + \tau q_-')   + (q_- + \tau q_-')\ln\left(\frac{q_- + \tau q_-'}{v_L}\right) \right] \, \rmd\tau  \right\}.
\end{eqnarray}
where the integral is to be minimized over all $q_+(\tau)$, $q_-(\tau)$ obeying the boundary condition $q_-(t_0)=j_{-0}$, $q_+(t_0)=j_{+0}$, $q_-(t)=j_{-}$, $q_+(t)=j_{+}=j+j_{-}$.  Here one has to solve two coupled Euler-Lagrange equations
\begin{equation}
\begin{array}{ll}
\fl \frac{\partial v_R}{\partial q_+} - q_+ \frac{\partial v_R / \partial q_+}{v_R} + \tau q_-'\frac{\partial v_R / \partial q_-}{v_R} + \frac{\partial{v_L}}{\partial q_+} - (q_-+\tau q_-')\frac{\partial v_L / \partial q'_+}{v_L} - \frac{2 \tau q'_+}{q_+ +\tau q'_+} - \frac{\tau^2 q''_+}{q_+ +\tau q'_+}&=0 \\
\fl \frac{\partial v_L}{\partial q_-} - q_- \frac{\partial v_L / \partial q_-}{v_L} + \tau q_+'\frac{\partial v_L / \partial q_+}{v_L} + \frac{\partial{v_R}}{\partial q_-} - (q_+ +\tau q_+')\frac{\partial v_R / \partial q'_-}{v_R} - \frac{2 \tau q'_-}{q_- +\tau q'_-} - \frac{\tau^2 q''_-}{q_- +\tau q'_-}&=0.
\end{array}
\end{equation}
These equations are exactly solvable for various choices of memory dependence including the interesting case of ``activity''-dependent rates with a constant bias which we explore in the next paragraph.
 
\begin{figure}
\begin{center}
\psfrag{p1}[Tc][Bc]{\footnotesize{$-\ln(P)/t^{1-a-c}$}}
\psfrag{p}[Tc][Bc]{$-\ln(P)/t$}
\psfrag{j}[Tc][Bc]{$j$}
\psfrag{j1}[Tc][Bc]{\footnotesize{$j$}}
\includegraphics{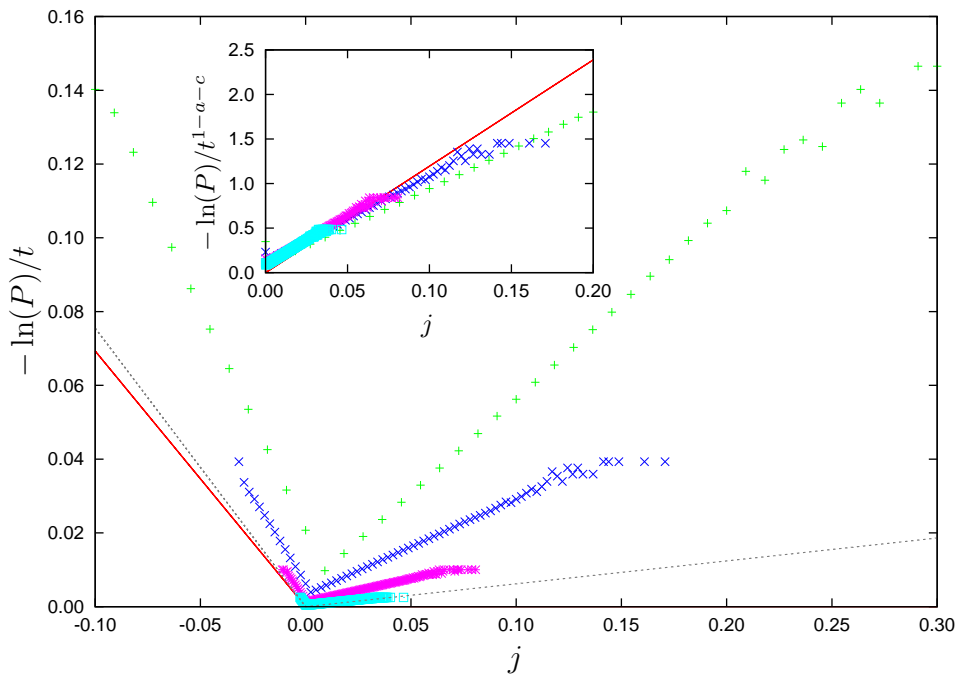}
\caption{Simulation results for distribution of currents in activity-dependent random walk with $a=0.4$, $c=0.2$, $t_0=10$, $j_{+0}=0.1$, $j_{-0}=0$.  Main plot shows $-\ln[\text{Prob}(\mathcal{Q}_t/t=j)]/t$ versus $j$: points are simulation data for $t=110,410,1610,6410$ (top to bottom); dotted line is prediction of~(\ref{e:act}) for $t=6410$ and solid line the infinite time limit.  Inset shows $-\ln[\text{Prob}(\mathcal{Q}_t/t=j)]/t^{1-a-c}$ for positive $j$ (same parameters).} 
\label{f:biack}
\end{center}
\end{figure}

A measure of the total activity of the system is given by the symmetric sum of left and right currents; for our second concrete example we consider rates proportional to the time average of this quantity, viz.\ $v_R(q_+,q_-)=a (q_+ + q_-)$, $v_L(q_+,q_-)= c (q_+ + q_-)$.   Without loss of generality, we take $a > c$, i.e., a drive to the right.  For the case $a+c<1$, after solving the Euler-Lagrange equations with the required boundary conditions, carrying out the integration and minimizing with respect to $j_{-}$ one obtains the leading terms in the exponent of the asymptotic distribution
\begin{equation}
\text{Prob}(\mathcal{Q}_t/t=j) \sim
\left\{
\begin{array}{ll}
\exp[ -j t_0^{a+c} \left(\frac{a+c}{a-c}\right) t^{1-a-c}] & \text{for $j \geq 0$} \\
\exp[ j (\ln\frac{a}{c}) t + j t_0^{a+c} \left(\frac{a+c}{a-c}\right) t^{1-a-c} ] \qquad & \text{for $j < 0$}.
\end{array}
\right. \label{e:act}
\end{equation}
Note that for $c=0$ this reduces to~(\ref{e:exp}) as it should.  
The form of this distribution is in good agreement with the output of numerical simulation as shown in figure~\ref{f:biack}.  
In particular, we observe that the dominant term in the exponent is different for positive currents than for negative currents --- being proportional to respectively $t^{1-a-c}$ and $t$.    In a strict mathematical sense this means that $\mathcal{Q}_t/t$ obeys a large deviation principle with speed $t$ and rate function which is only non-zero for $j < 0$.   Physically, it implies different scaling for fluctuations in the forward and backward directions.

Even for single-particle random walks there are relatively few choices for the rate-dependence which lead to Euler-Lagrange equations susceptible to exact solution.  However, 
in many other cases one can make progress via a Gaussian approximation analogous to that applied for the spatial additivity principle~\cite{Bodineau04}.  For instance, for the bi-directional random walk one makes an expansion about the mean steady-state currents $\bar{q}_+(\tau)$ and $\bar{q}_-(\tau)$ which obey $\bar{q}_+ = v_R(\bar{q}_+,\bar{q}_-)$ and $\bar{q}_- = v_L(\bar{q}_+,\bar{q}_-)$.  Such an approach allows us to treat, for example, the case $v_R(q_+,q_-)=a q_+ + b, v_L(q_+,q_-)=cq_+ + d$ and obtain results which are essentially just asymmetric generalizations of the original discrete-time elephant in~\cite{Schutz04}.  For example, for $0<a,c<1/2$ one finds
\begin{equation}
\text{Prob}(\mathcal{Q}_t/t=j) \sim \exp\left\{ -\left[ \frac{1}{2}\frac{ \left( j - \frac{b}{1-a} + \frac{d}{1-c}\right)^2}{\frac{b}{(1-a)(1-2a)}+\frac{d}{(1-c)(1-2c)}} \right] t \right\}. \label{e:eleph1}
\end{equation}
For $1/2<a,c<1$ one again has ``superdiffusive'' behaviour, e.g., for $a=c$
\begin{equation}
\text{Prob}(\mathcal{Q}_t/t=j) \sim \exp\left\{ -\left[ \frac{1}{2}\frac{ \left( j - \frac{b}{1-a} + \frac{d}{1-a}\right)^2}{\frac{b+d}{(1-a)(2a-1)}} \right] t_0^{2a-1} t^{2-2a} \right\}.  \label{e:eleph2}
\end{equation}

\section{Fluctuation symmetries}

Here we use the bi-directional examples of the previous section to illustrate the validity of fluctuation symmetries for memory-dependent systems.  First, we note that for the random walk with activity-dependent rates~(\ref{e:act}) imposes
\begin{equation}
\frac{\text{Prob}(\mathcal{Q}_t/t=-j)}{\text{Prob}(\mathcal{Q}_t/t=+j)} \sim \exp \left[ -j (\ln \frac{a}{c}) t \right]
\end{equation}
which is just the usual Gallavotti-Cohen fluctuation relation~\cite{Evans93,Gallavotti95,Lebowitz99}.  Indeed this is to be expected since, for this particular model, the relative bias felt by the particle is constant, $v_R/v_L=a/c$.\footnote{In fact, it seems that the system also shows the symmetry for the case $a+c>1$ when there is obviously no stationary state in the usual sense.  This is a simple counterexample to the suggestion in~\cite{Speck07b} that the existence of a stationary state is necessary for fluctuation relations in non-Markovian processes.}
In a sense, this example is thus rather trivial but it is already an illuminating reminder that the validity of the fluctuation theorem is not restricted to the usual Markovian cases where one has $\text{Prob}(\mathcal{Q}_t/t=j) \sim \rme^{-tI(j)}$ for both positive and negative currents.

Now we turn to consider a biased elephant walker with $v_R(q_+,q_-)=a q_+ + b$, $v_L(q_+,q_-)=aq_+ + d$.  Within the Gaussian approximation given by~(\ref{e:eleph1}) and~(\ref{e:eleph2}), we have
\begin{equation}
\frac{\text{Prob}(\mathcal{Q}_t/t=-j)}{\text{Prob}(\mathcal{Q}_t/t=+j)} \sim 
\left\{
\begin{array}{ll}
\exp \left[ -j \frac{2(b-d)(1-2a)}{1-a} t \right] & \text{for $0<a<1/2$} \\
\exp \left[ -j \frac{2(b-d)(1-2a)}{1-a} t_0^{2a-1} t^{2-2a} \right] & \text{for $1/2<a<1$.}
\end{array}
\right.
\end{equation}
We remark that in both cases there is a well-defined mean current given by $\bar{q}=(b-d)/(1-a)$ but only for $0<a<1/2$ is a fluctuation relation of the usual form recovered.  For $1/2<a<1$ the symmetry is apparently modified by the ``superdiffusive'' spreading about the mean. In this case, the logarithm of the ratio of probabilities for currents $-j$ and $+j$ is still asymptotically proportional to $j$ but is sublinear in $t$. This scenario should merit closer investigation in other models.

\section{Discussion and further applications}

In principle, the temporal additivity principle presented above can be applied to any stochastic system where the corresponding (Markovian) large deviation function $I_w(j)$ is known.  In practice, the resulting Euler-Lagrange equation may not be analytically tractable.  However, we suggest that the approach still has considerable numerical utility since the integral in~(\ref{e:result}) should be amenable to minimization by standard algorithms.  In particular, it would be interesting to investigate the applicability of the method to systems with dynamical phase transitions.  One such model is the zero-range process with open boundaries in which exact expressions for the Markovian large deviation functions are known and can be related to those of random walks with certain effective rates~\cite{Rakos08}.  This suggests that some of the memory-induced effects observed in the simple examples of section~\ref{s:rw} may also be observed in more complicated many-particle systems.  In particular, we conjecture that long-range temporal correlations in non-equilibrium systems may alter the speed of the current large deviation function in an analogous way to long-range spatial correlations modify the speed of equilibrium large deviations~\cite{Cox85,Ellis95}. Further work in this area is needed to establish the exact link between the speed of a large deviation principle and the ``long-rangedness'' of correlations.

Finally, we remark that the additivity formalism could also be applied to intrinsically non-Markovian systems where the rates at time $\tau$ depend, e.g., on $\mathcal{Q}_{\tau'}/{\tau'}$ where $\tau'=x\tau$ and $x$ is a positive constant less than unity (cf.\ the ``Alzheimer'' random walk~\cite{daSilva05,Cressoni07,Kenkre07}).  However, in that case, the integrand in~(\ref{e:result}) becomes non-local and minimization correspondingly difficult.  There is clearly much scope for future analytical work on current fluctuations in non-Markovian systems as well as for the development of efficient simulation algorithms for measuring the corresponding large deviation functions.  

Note added: After submission of this work there appeared an interesting preprint~\cite{Maes09b} discussing similar issues for a different class of non-Markovian systems, namely semi-Markov processes.

\section*{References}

\bibliographystyle{myunsrtb}
%\bibliography{/home/rosemary/allref}
\bibliography{/home/network/harris/allref}

\begin{thebibliography}{10}

\bibitem{Derrida07b}
B.~Derrida.
\newblock Non-equilibrium steady states: fluctuations and large deviations of
  the density and of the current.
\newblock {\em J. Stat. Mech.},  P07023, 2007.

\bibitem{Touchette09b}
H.~Touchette.
\newblock The large deviation approach to statistical mechanics.
\newblock To be published in Phys. Rep., DOI: 10.1016/j.physrep.2009.05.002,
  2009.

\bibitem{Evans93}
D.~J. Evans, E.~G.~D. Cohen, and G.~P. Morriss.
\newblock Probability of second law violations in shearing steady states.
\newblock {\em Phys. Rev. Lett.}, 71(15):2401--2404, 1993.

\bibitem{Gallavotti95}
G.~Gallavotti and E.~G.~D. Cohen.
\newblock Dynamical ensembles in nonequilibrium statistical mechanics.
\newblock {\em Phys. Rev. Lett.}, 74(14):2694--2697, 1995.

\bibitem{Lebowitz99}
J.~L. Lebowitz and H.~Spohn.
\newblock A {G}allavotti-{C}ohen-type symmetry in the large deviation
  functional for stochastic dynamic.
\newblock {\em J. Stat. Phys.}, 95(1--2):333--365, 1999.

\bibitem{Me07}
R.~J. Harris and G.~M. Sch\"utz.
\newblock Fluctuation theorems for stochastic dynamics.
\newblock {\em J. Stat. Mech.},  P07020, 2007.

\bibitem{Rangarajan03}
G.~Rangarajan and M.~Ding, editors.
\newblock {\em Processes with Long-Range Correlations: Theory and
  Applications}, volume 621 of {\em Lecture Notes in Physics}.
\newblock Springer, Berlin, 2003.

\bibitem{Mandelbrot99}
B.~B. Mandelbrot.
\newblock {\em Multifractals and $1/f$ Noise: Wild Self-Affinity in Physics}.
\newblock Springer, New York, 1999.

\bibitem{benAvraham00}
D.~{ben-Avraham} and S.~Havlin.
\newblock {\em Diffusion and Reactions in Fractals and Disordered Systems}.
\newblock Cambridge University Press, Cambridge, 2000.

\bibitem{Mantegna99}
R.~N. Mantegna and H.~E. Stanley.
\newblock {\em Introduction to Econophysics: Correlations and Complexity in
  Finance}.
\newblock Cambridge University Press, Cambridge, 1999.

\bibitem{Hanggi78}
P.~H\"anggi.
\newblock Correlation functions and masterequations of generalized
  (non-{M}arkovian) {L}angevin equations.
\newblock {\em Z. Phys. B}, 31(4):407--416, 1978.

\bibitem{Volkov83}
V.~S. Volkov and V.~N. Pokrovsky.
\newblock Generalized {F}okker-{P}lanck equation for non-{M}arkovian processes.
\newblock {\em J. Math. Phys.}, 24(2):267--270, 1983.

\bibitem{Metzler99}
R.~Metzler, E.~Barkai, and J.~Klafter.
\newblock Anomalous diffusion and relaxation close to thermal equilibrium: A
  fractional {F}okker-{P}lanck equation approach.
\newblock {\em Phys. Rev. Lett.}, 82(18):3563--3567, 1999.

\bibitem{Montroll65}
E.~W. Montroll and G.~H. Weiss.
\newblock Random walks on lattices {II}.
\newblock {\em J. Math. Phys.}, 6(2):167--181, 1965.

\bibitem{Murray02}
J.~D. Murray.
\newblock {\em Mathematical Biology I: An Introduction}.
\newblock Springer, New York, 3rd edition, 2002.

\bibitem{Mai07}
T.~Mai and A.~Dhar.
\newblock Nonequilibrium work fluctuations for oscillators in non-{M}arkovian
  baths.
\newblock {\em Phys. Rev. E}, 75(6):061101, 2007.

\bibitem{Speck07b}
T.~Speck and U.~Seifert.
\newblock The {J}arzynski relation, fluctuation theorems, and stochastic
  thermodynamics for non-{M}arkovian processes.
\newblock {\em J. Stat. Mech.},  L09002, 2007.

\bibitem{Ohkuma07}
T.~Ohkuma and T.~Ohta.
\newblock Fluctuation theorems for non-linear generalized {L}angevin systems.
\newblock {\em J. Stat. Mech.},  P10010, 2007.

\bibitem{Esposito08}
M.~Esposito and K.~Lindenberg.
\newblock Continuous-time random walk for open systems: Fluctuation theorems
  and counting statistics.
\newblock {\em Phys. Rev. E}, 77(5):051119, 2008.

\bibitem{Andrieux08d}
D.~Andrieux and P.~Gaspard.
\newblock The fluctuation theorem for currents in semi-{M}arkov processes.
\newblock {\em J. Stat. Mech.},  P11007, 2008.

\bibitem{Chechkin09}
V.~Chechkin and R.~Klages.
\newblock Fluctuation relations for anomalous dynamics.
\newblock {\em J. Stat. Mech.},  L03002, 2009.

\bibitem{Hod04}
S.~Hod and U.~Keshet.
\newblock Phase transition in random walks with long-range correlations.
\newblock {\em Phys. Rev. E}, 70(1):015104, 2004.

\bibitem{Schutz04}
G.~M. Sch\"utz and S.~Trimper.
\newblock Elephants can always remember: Exact long-range memory effects in a
  non-{M}arkovian random walk.
\newblock {\em Phys. Rev. E}, 70:045101, 2004.

\bibitem{daSilva05}
M.~A.~A. da~Silva, J.~Cressoni, and G.~M. Viswanathan.
\newblock Discrete-time non-{M}arkovian random walks: The effect of memory
  limitations on scaling.
\newblock {\em Physica A}, 364:70--78, 2006.

\bibitem{Paraan06}
F.~N.~C. Paraan and J.~P. Esguerra.
\newblock Exact moments in a continuous time random walk with complete memory
  of its history.
\newblock {\em Phys. Rev. E}, 74(3):032101, 2006.

\bibitem{Cressoni07}
J.~C. Cressoni, M.~A.~A. da~Silva, and G.~M. Viswanathan.
\newblock Amnestically induced persistence in random walks.
\newblock {\em Phys. Rev. Lett.}, 98(7):070603, 2007.

\bibitem{Kenkre07}
V.~M. Kenkre.
\newblock Analytic formulation, exact solutions, and generalizations of the
  elephant and the {A}lzheimer random walks.
\newblock Preprint arXiv:0708.0034, 2007.

\bibitem{Cleuren03}
B.~Cleuren and C.~{Van den Broeck}.
\newblock Brownian motion with absolute negative mobility.
\newblock {\em Phys. Rev. E}, 67(5):055101, 2003.

\bibitem{Bodineau04}
T.~Bodineau and B.~Derrida.
\newblock Current fluctuations in non-equilibrium diffusive systems: an
  additivity principle.
\newblock {\em Phys. Rev. Lett.}, 92(18):180601, 2004.

\bibitem{Ellis95}
R.~S. Ellis.
\newblock An overview of the theory of large deviations and applications to
  statistical mechanics.
\newblock {\em Scand. Actuarial J.}, 1(1):97--142, 1995.

\bibitem{Derrida04b}
B.~Derrida, B.~Doucot, and P.-E. Roche.
\newblock Current fluctuations in the one dimensional symmetric exclusion
  process with open boundaries.
\newblock {\em J. Stat. Phys.}, 115(3--4):717--748, 2004.

\bibitem{Bodineau06}
T.~Bodineau and B.~Derrida.
\newblock Current large deviations for asymmetric exclusion processes with open
  boundaries.
\newblock {\em J. Stat. Phys.}, 123(2):277--300, 2006.

\bibitem{Wijland04}
F.~van Wijland and Z.~R{\'a}cz.
\newblock Large deviations in weakly interacting boundary driven lattice gases.
\newblock {\em J. Stat. Phys.}, 118(1--2):27--54, 2005.

\bibitem{Me05}
R.~J. Harris, A.~R{\'a}kos, and G.~M. Sch{\"u}tz.
\newblock Current fluctuations in the zero-range process with open boundaries.
\newblock {\em J. Stat. Mech.},  P08003, 2005.

\bibitem{Me06b}
R.~J. Harris, A.~R{\'a}kos, and G.~M. Sch{\"u}tz.
\newblock Breakdown of {G}allavotti-{C}ohen fluctuation theorem for stochastic
  dynamics.
\newblock {\em Europhys. Lett.}, 75(2):227--233, 2006.

\bibitem{Dembo98}
A.~Dembo and O.~Zeitouni.
\newblock {\em Large Deviation Techniques and Applications}.
\newblock Springer, New York, 2nd edition, 1998.

\bibitem{Rakos08}
A.~R\'akos and R.~J. Harris.
\newblock On the range of validity of the fluctuation theorem for stochastic
  {M}arkovian dynamics.
\newblock {\em J. Stat. Mech.},  P05005, 2008.

\bibitem{Cox85}
J.~T. Cox and D.~Griffeath.
\newblock Large deviations for some infinite particle system occupation times.
\newblock {\em Contemporary Math.}, 41:43--54, 1985.

\bibitem{Maes09b}
C.~Maes, K.~Neto\v{c}n\'y, and B.~Wynants.
\newblock Dynamical fluctuations for semi-{M}arkov processes.
\newblock Preprint arXiv:0905.4897, 2009.

\end{thebibliography}

\end{document}